\documentclass[aps,prl,twocolumn,superscriptaddress,showpacs]{revtex4}
\usepackage{graphicx}
\usepackage{mathrsfs}
\usepackage{bm}

\begin{document}

\title{Quantum Anomalous Hall Effect in Graphene Proximity-Coupled to an Antiferromagnetic Insulator}
\author{Zhenhua Qiao}
\affiliation{Department of Physics, University of Science and Technology of China, Hefei, Anhui 230026, China}
\affiliation{ICQD, Hefei National Laboratory for Physical Sciences at Microscale, University of Science and Technology of China, Hefei, Anhui 230026, China}
\affiliation{Department of Physics, The University of Texas at Austin, Austin, Texas 78712, USA}
\author{Wei Ren}
\affiliation{Department of Physics, Shanghai University, Shanghai 200444, China}
\affiliation{Department of Physics, The University of Arkansas, Fayetteville, Arkansas 72701, USA}
\author{Hua Chen}
\affiliation{Department of Physics, The University of Texas at Austin, Austin, Texas 78712, USA}
\author{L. Bellaiche}
\affiliation{Department of Physics, The University of Arkansas, Fayetteville, Arkansas 72701, USA}
\author{Zhenyu Zhang}
\affiliation{Department of Physics, The University of Texas at Austin, Austin, Texas 78712, USA}\affiliation{ICQD, Hefei National Laboratory for Physical Sciences at Microscale, University of Science and Technology of China, Hefei, Anhui 230026, China}
\author{A.H. MacDonald}
\affiliation{Department of Physics, The University of Texas at Austin, Austin, Texas 78712,
USA}
\author{Qian Niu}
\affiliation{Department of Physics, The University of Texas at Austin, Austin, Texas 78712,
USA}\affiliation{International Center for Quantum Materials, Peking University, Beijing 100871, China}

\begin{abstract}
  We propose realizing the quantum anomalous Hall effect by proximity coupling graphene to an antiferromagnetic insulator that provides both broken time-reversal symmetry and spin-orbit coupling. We illustrate our idea by performing {\em ab initio} calculations for graphene adsorbed on the (111) surface of BiFeO$_3$.  In this case we find that the proximity-induced exchange field in graphene is 70 meV, and that a topologically nontrivial band gap is opened by Rashba spin-orbit coupling. The size of the gap depends on the separation between the graphene and the thin film substrate, which can be tuned experimentally by applying external pressure.
\end{abstract}
\pacs{
71.70.Ej,  
73.43.Cd,  
81.05.Uw   
}
\maketitle

The possibility of Hall conductance quantization in a two-dimensional
electron systems in the absence of an external magnetic field was first proposed by Haldane
based on honeycomb lattice~\cite{haldane} toy model studies.
Like the conventional quantum Hall effect, this quantum {\em anomalous} Hall effect (QAHE)
can occur only in two-dimensional systems that are not time-reversal invariant.
Although its appearance is dependent on electronic structure details,
the QAHE is a possibility in any strongly spin-orbit coupled
magnetic two-dimensional electron systems.  The QAHE has so far been predicted
to occur in mercury-based quantum wells~\cite{SCZhang1},
topological insulator thin films~\cite{SCZhang3,JiangHua,Organic},
graphene~\cite{qiao1,Hongbin,James} and silicene~\cite{Ezawa} based systems,
kagome lattice~\cite{kagome} systems,
and in thin layers containing heavy elements~\cite{Vanderbilt}.
Related effects can occur in optical lattices \cite{Congjun,yongping}.
The QAHE has recently been observed experimentally in
chromium-doped (Bi,Sb)$_2$Te$_3$~\cite{Chang}.

Graphene is unique among two-dimensional electron systems in its ability to maintain electrical isolation
at high carrier densities, and is in many ways an ideal system for applications of the
QAHE.   However graphene is not magnetic in its pristine form and
has extremely weak intrinsic spin-orbit coupling~\cite{WeakSOC1,WeakSOC2,WeakSOC3}.
In order to realize the QAHE effect both spin-orbit coupling and magnetism
should be induced externally, for example by depositing 3$d$ transition metal atoms on graphene~\cite{qiao3}.
However, metal adatoms tend to nucleate into clusters on graphene~\cite{clustering,Chenhua}. As reported in Refs.~\cite{magneticTI}, ferromagnetism in topological insulators can be induced due to the proximity coupling to ferromagnetic insulators. Comparing to the spreading of surface states of a topological insulator into the thickness direction of a few quintuple layers, the single atomic-layer graphene gives a much stronger hybridization with the substrate, therefore the proximity effect in graphene should be generally larger than that in topological insulators. In this Letter we explore the possibility of depositing graphene on a
proper magnetic substrate, which can induce both a time-reversal symmetry breaking
exchange field and Rashba spin-orbit coupling due to broken inversion symmetry.
Guided by previous work\cite{qiao1} we expect the QAHE to be realized when the substrate
satisfies the following criteria:
(i) it is insulating so that graphene provides the only transport channel; (ii)
its orbitals hybridize with those of graphene sufficiently to induce sizable exchange and
spin-orbit coupling fields; (iii)
its dipolar magnetic field is sufficiently weak that the quantum anomalous Hall signature
is not obscured by normal quantum Hall effect.
In this Letter we demonstrate that antiferromagnetic insulators with properly terminated surfaces
can satisfy all these criteria.
BiFeO$_3$ is a heavily studied room-temperature multiferroic material
(see, e.g. Ref.~\cite{albrecht} and references therein).
By theoretically depositing graphene on the (111) surface, we
show that the graphene $\pi$ bands spin-split on a BiFeO$_3$ thin film substrate.
After switching on spin-orbit coupling in our electronic structure calculations,
we further find that a small band gap $\delta$ opens near the graphene K and K'
Dirac points which are located inside the bulk band gap $\Delta$ of the BiFeO$_3$ thin film substrate.
Berry curvature analysis confirms that the resulting band gap is topologically nontrivial,
yielding a two-dimensional quantum anomalous Hall phase.
Finally, we show that $\delta$ can be increased to a value that is large enough
for high temperature operation by reducing the
separation between graphene and the thin film substrate.

Our first-principles calculations were performed using the projected-augmented-wave method~\cite{PAW}
as implemented in the Vienna Ab-initio Simulation Package (VASP)~\cite{VASP}.
The generalized gradient approximation (GGA)~\cite{GGA} exchange-correlation functional was adopted.
The value of the Hubbard U was chosen to be 3~eV to account for strong intra-atomic interactions on the Fe atoms
and to correct band gap underestimation. The lattice constant of graphene
was chosen to be $a=2.39{\rm \AA}$, and the kinetic energy cutoff was set to be 550 eV.
During structural relaxation, all atoms were allowed to relax along any direction, and forces
were converged to less than 0.01~$\rm eV/{ \AA}$.
The first Brillouin-zone integration was carried out using $3 \times 3 \times1$ Monkhorst-Pack grids.
A vacuum buffer space over 15$\rm {\AA}$ was included to prevent interaction between adjacent slabs.
Hydrogen atoms were used to passivate the Bi surface on the side of the BiFeO$_3$ film
opposite to the graphene interface.

\begin{figure}
  \includegraphics[width=8cm,angle=0]{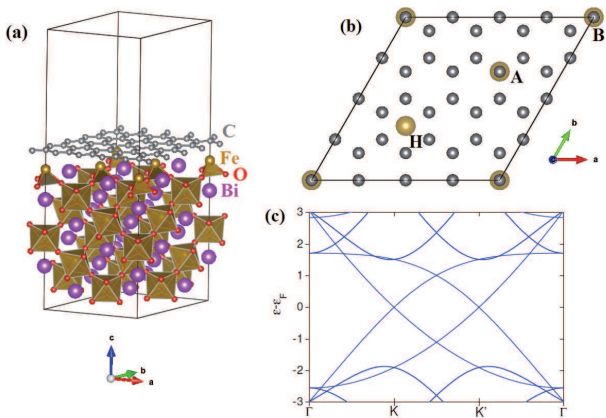}
  \caption{(Color online) (a) Supercell for graphene on a (111) BiFeO$_3$ surface. Gray, purple, yellow, and red colors represent respectively C, Bi, Fe, and O atoms. (b) A top view of graphene with one layer of iron atoms. One Fe atom is located at the hollow site labeled by H, while the other two Fe atoms are located at the top sites labeled as A and B. (c) Band structure along high symmetry lines for a $4\times4$ supercell of graphene.
  } \label{SupercellBFO-Gra}
\end{figure}

BiFeO$_3$ has a perovskite structure in which the
magnetic Fe ions form a simple cubic lattice.
Since BiFeO$_3$ is a G-type antiferromagnet, the Fe moments in each (111) plane are aligned
and neighboring (111) planes have opposite spin polarizations.
We therefore expect that a (111) BiFeO$_3$ surface, which hybridizes relatively strongly
with a graphene sheet deposited on it, can induce both a strong homogeneous exchange field and
observable spin-orbit coupling. Because the overall order in BiFeO$_3$
is antiferromagnetic, it will not induce significant
magnetostatic magnetic fields in graphene.
Figure~\ref{SupercellBFO-Gra} (a) illustrates the supercell used in our calculations,
which consists of a graphene 4$\times$4 unit cell on a 4-layer BiFeO$_3$ slab with a
(111) surface.
The lattice mismatch between graphene and BiFeO$_3$ in this arrangement is less than $2\%$.
The average distance between graphene and the Fe layer after relaxation is found to be $d\approx$ 2.60\AA.
As illustrated in Fig.~\ref{SupercellBFO-Gra} (b) there are three Fe atoms per unit cell in the surface layer:
one located at the hollow site labeled H and the other two are located at the top sites labeled A and B.
Since the two top sites belong to different sublattices of graphene, the surface
structure helps to suppress induced staggered sublattice potentials, which
favors valley-Hall-effect gaps in graphene~\cite{QVHE} and are detrimental to QAHE.

\begin{figure*}
\includegraphics[width=14cm,angle=0]{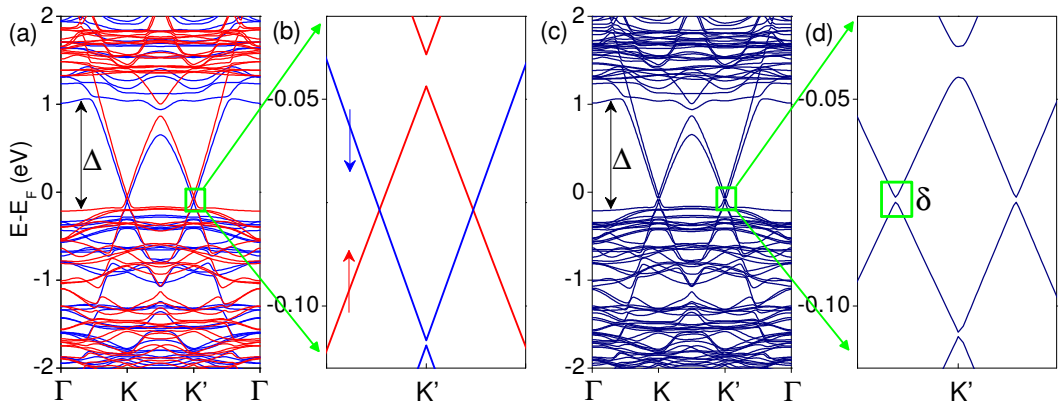}
\caption{(Color online) Bulk band structure of graphene on BiFeO$_3$. (a) Spin-resolved band structures. (b) Zoom-in on the green box of panel (a). The blue and red in panels (a) and (b) plot the spin-up and spin-down bands, respectively. (c) Band structure when spin-orbit coupling included. A bulk gap $\delta$ of about 1.1 meV is centered on the avoided
crossings near the K and K' valleys.
Note that the nontrivial band gap $\delta$ is inside the bulk gap $\Delta$ of BiFeO$_3$.
(d) Zoom in on the green box of panel (c).} \label{BFO-bands}
\end{figure*}

Figure~\ref{SupercellBFO-Gra}(c) plots the band structure of an isolated graphene
layer along the high symmetry lines of the $4\times4$ supercell Brillouin zone
used in our supercell calculations.
When graphene is placed on top of the BiFeO$_3$ substrate, it is
magnetized by the exchange interaction between its $\pi$-orbitals
and the 3$d$ orbitals of the ferromagnetic Fe layer, which is antiferromagnetic relative to the surface of BiFeO$_3$ layer.
Figure~\ref{BFO-bands} (a) illustrates the spin-polarized band structure of graphene
on the (111) surface of BiFeO$_3$ thin film substrate.
Figure~\ref{BFO-bands} (b) zooms in the spin-polarized $\pi$-bands
near the Dirac point at K' and shows that the induced exchange splitting is
$M \sim 70$~meV. 
Note that there is a small Dirac point splitting between bands of the same spin
which can be attributed to a small non-zero staggered AB sublattice potential in the relaxed structure.
We expect that the staggered sub lattice potential will be highly random
in realistic systems and therefore do not expect that this splitting will play an important role.
Another important observation from Fig.~\ref{BFO-bands} (a) is that the Dirac points highlighted by the green box
are located inside the bulk band gap of the thin film substrate.

\begin{figure}
\includegraphics[width=7cm,angle=0]{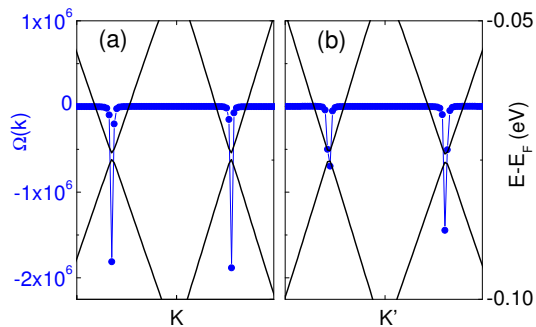}
\caption{(Color online) Berry curvature along high symmetry lines summed over occupied valence bands below.}\label{BerryPhase}
\end{figure}

When spin-orbit coupling is included a small band gap opens
at the avoided crossing between crossing bands with opposite spin-orientations, as illustrated
in Figs.~\ref{BFO-bands} (c) and (d). Figure \ref{BFO-bands}
(d) zooms in the bands of Fig.~\ref{BFO-bands}(c) near the Dirac point K' and shows that the avoided crossing
gap $\delta\approx 1.1$ meV.   Experimentally the Fermi level $E_F$ can be tuned to lie in this gap by a back gate under
the substrate.  To determine whether or not the resulting insulating state exhibits the QAHE
we integrate momentum-space Berry curvatures over the supercell Brillouin-zone.
Figure~\ref{BerryPhase} plots the Berry curvature $\Omega(\bm k)$ summed over occupied valence bands:
\begin{equation}
\Omega (\bm k)=-\sum_n f_n \sum_{n'\neq n } \frac{2{\rm Im} \langle \psi_{n \bm k}|v_x| \psi_{n' \bm k}\rangle \langle \psi_{n' \bm k} |v_y|\psi_{n \bm k}\rangle}{(E_{n'}-E_n)^2},
\end{equation}
along high symmetry lines.
Here $v_x$ and $v_y$ are operator components along the $x$ and $y$ directions and $f_n=1$ for
occupied bands.
One can see that the Berry curvature is non-zero and that it has the same sign
near the K and K' valley points.
As a consequence its integration over the Brillouin zone must give rise to a nonzero Chern number $\mathcal{C}$:
\begin{equation}
\mathcal{C}=\frac{1}{2\pi}\int_{BZ} d^2 k~\Omega (\bm k).
\end{equation}
The quantized anomalous Hall conductance $\sigma_{xy}$ is related to the Chern number $\mathcal{C}$ by $\sigma_{xy}=\mathcal{C} e^2/h$. Due to the absence of external magnetic field, our obtained insulating state is the long-sought quantum anomalous Hall effect.

According to Ref.~[\onlinecite{qiao1}], both engineered exchange-coupling and
engineered Rashba spin-orbit coupling must play a crucial role in any graphene based QAHE.
In the present case, both terms arise from the interaction between graphene and the BiFeO$_3$ thin film substrate.
The former originates from the proximity exchange coupling and the latter by breaking the mirror symmetry
of the graphene plane.  A staggered sublattice potential can also influence the
quantum anomalous Hall gap. The low-energy continuum model of graphene when
all of these effects are present is:~\cite{James,qiao3}
\begin{eqnarray}
H({\bm k})=&-&v_f (\eta \sigma_x k_x + \sigma_y k_y){\bm 1}_s+M {\bm 1}_\sigma s_z \nonumber \\
&+&\frac{\lambda_R}{2}(\eta \sigma_x s_y -\sigma_y s_x) + U \sigma_z {\bm 1}_s,
\label{Hamiltonian}
\end{eqnarray}
where $v_f=3t/2$ is the Fermi velocity, $t \sim 2.60$ eV, $\eta=\pm 1$ for $K,K'$ respectively, and
$\bm \sigma$ and $\bm s$ are Pauli matrices that act on sublattice and spin degrees of freedom.
The first term in Eq.~\ref{Hamiltonian} is the Hamiltonian of pristine graphene, the second describes
$\pi$-electron exchange coupling to the magnetic order, the third is Rashba spin-orbit coupling
and the final term is the staggered sub lattice potential.
By fitting Eq.~\ref{Hamiltonian} to the first-principles band structure near the Dirac points, we can extract values of the three
coupling parameters.  
For example, Rashba spin-orbit coupling in the present system is estimated to
have coupling constant $\lambda_R\approx 1.26$ meV.

\begin{figure}
\includegraphics[width=7cm,angle=0]{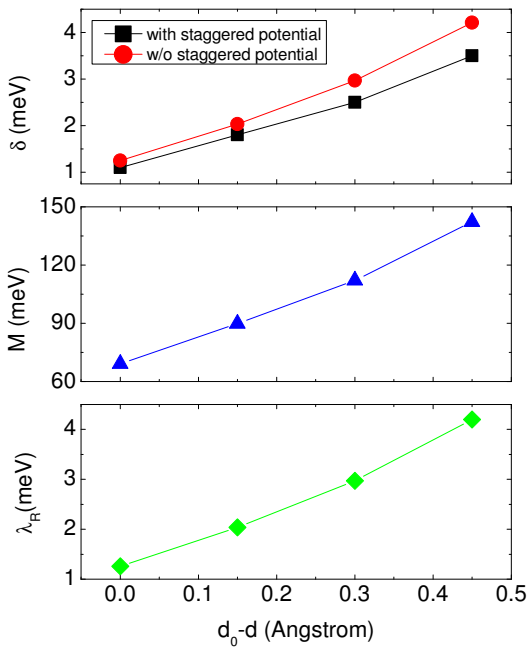}
\caption{(Color online) (a) Bulk band gap $\delta$ as a function of a
change $(d_0-d)$ in the separation between the graphene sheet and the BiFeO$_3$ ($d_0$ is the optimal distance obtained from the full relaxation). The square and circle symbols refer respectively to gaps from density
{\em ab initio} calculations and gaps from the low-energy model
with the staggered sublattice potential set to zero.
(b) Exchange field $M$ as a function of $(d_0-d)$.
(c) Rashba spin-orbit coupling strength as a function of $(d_0-d)$.} \label{Gap-d-BFO}
\end{figure}

Since both exchange field and Rashba spin-orbit coupling are induced by hybridization between graphene and thin film substrate orbitals,
we expect that they will strengthen if the separation between the graphene sheet and the thin film substrate is
reduced.  In experiment this can be achieved by applying an external uniaxial stress perpendicular to the graphene sheet.
In Fig.~\ref{Gap-d-BFO}, we plot the band gap $\delta$, and the
exchange field $M$ and Rashba spin-orbit coupling $\lambda_R$ strengths
{\em vs.} separation change $d_0-d$.  We find that $\lambda_R$ increases by a factor of $\sim 3$,
$M$ by a factor of $\sim 2$, and $\delta$ by a factor of $\sim 3$ when the separation is reduced by $\sim 20$\%.
As expected, the band gap $\delta$ increases with decreasing $d$ [square symbols in Fig.~\ref{Gap-d-BFO}(a)]. When $d=2.15$~\AA, the band gap $\delta$ can reach 3.50 meV, corresponding to a temperature of $40.6$~K.
The low-energy model band gaps in Figure~\ref{Gap-d-BFO}(a) (red circles) do not
account for any potential influence of the staggered sublattice potential.
The close agreement demonstrates that this part of the Hamiltonian does not play a major
role in determining the size of the gap.

\begin{figure}
\includegraphics[width=6cm,angle=0]{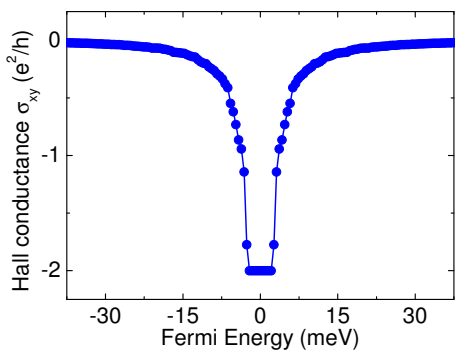}
\caption{(Color online) Anomalous Hall conductance as a function of Fermi level calculated from the tight-binding model Hamiltonian corresponding to Eq.(\ref{Hamiltonian}).} \label{HallConductance}
\end{figure}

In Figure~\ref{HallConductance} we plot the intrinsic\cite{Nagaosa_Review} anomalous Hall conductivity.
These results were calculated by integrating Berry curvatures calculated using a lattice representation of
Eq.~(\ref{Hamiltonian}) over the Brillouin-zone for $M=142$ meV and $\lambda_R=4.2$ meV, corresponding to
$(d_0-d)=0.45$~\AA.   As expected the Hall conductivity is quantized ($\sigma_{xy}=-2e^2/h$) when the Fermi level lies inside the bulk band gap.  We find that the intrinsic anomalous Hall conductivity rapidly approaches zero
when the Fermi level is outside the gap. This is expected since the weak Rashba spin-orbit coupling
implies that only bands near the band gap contribute to the nonzero Berry curvature~[see Fig.~\ref{BerryPhase}].

In summary, by performing explicit calculations for graphene on the (111) surface of BiFeO$_3$
we have demonstrated that the QAHE can be realized in graphene.  We expect that
our results apply to graphene on any uncompensated antiferromagnetic insulator surface.
Since many insulating materials have antiferromagnetic order at room temperature, this
effect raises the prospects for a high temperature QAHE which might have interesting potential
for applications.  In the BiFeO$_3$ case the topologically non-trivial gap is not sufficiently large to
achieve this goal, although it can be enhanced by applying pressure.
Our work motivates a search\cite{zeng} for other substrates.  For example, the (100) surfaces of A-type antiferromagnets are uncompensated and would work if the surface moments
have a nonzero component along the surface normal. When the van de Waals' interaction is further considered, the distance between graphene and the semi-infinite substrate is slightly decreased to $d_0=2.58\AA$, which naturally increases the resulting Rashba spin-orbit coupling and the QAHE gap.

\textit{Acknowledgements.---} The authors are grateful to Na Sai, Wanxiang Feng, Wenguang Zhu and Changgan Zeng for valuable discussions. This work was financially supported by the Welch Foundation (F-1255, TBF1473), DOE (DE-FG03-02ER45958, Division of Materials Science and Engineering) grant, NBRPC (2012CB-921300), NSFC (91021019, 91121004, 11274222, 11034006), Eastern Scholar Program, Shuguang Program (12SG34) from Shanghai Municipal Education Commission and Shanghai University Innovation Fund (SDCX2012027). Z.Q. also thanks the financial supporting by USTC Startup and 100 Talent Program of Chinese Academy of Sciences. W.R. and L.B. thank ARO Grant W911NF-12-1-0085 for personal support. They also acknowledge NSF DMR-1066158, ONR Grants N00014-11-1-0384 and N00014-12-1-1034, and Department of Energy, Office of Basic Energy Sciences, under contract ER-46612, for discussions with scientists sponsored by these grants. The Texas Advanced Computing Center (TACC) and Shanghai Supercomputer Center are gratefully acknowledged for high performance computing assistance.

\end{document}